\documentclass[useAMS,usenatbib,letterpaper]{mn2e}
\usepackage[totalwidth=480pt,totalheight=680pt]{geometry}
\include{graphicx}
\title[An optimal Earth Trojan asteroid search strategy]{An optimal Earth Trojan asteroid search strategy}

\author[M. Todd, P. Tanga, D. M. Coward and M. G. Zadnik]{M. Todd$^{1}$\thanks{E-mail:
michael.todd@icrar.org (MT)}, P. Tanga$^{2}$, D. M. Coward$^{3}$ and M. G. Zadnik$^{1}$\\
$^{1}$Department of Imaging and Applied Physics, Bldg 301, Curtin University, Kent St, Bentley, WA 6102, Australia\\
$^{2}$Laboratoire Cassiop\'{e}e, Observatoire de la C\^{o}te d'Azur, BP 4229, 06304 Nice Cedex 04, France\\
$^{3}$School of Physics, M013, The University of Western Australia, 35 Stirling Hwy, Crawley, WA 6009, Australia}

\begin{document}

\date{Accepted 2011 November 3.  Received 2011 November 3; in original form 2011 August 1}

\pagerange{\pageref{firstpage}--\pageref{lastpage}} \pubyear{2011}

\maketitle

\label{firstpage}

\begin{abstract}
Trojan asteroids are minor planets that share the orbit of a planet about the Sun and librate around the L4 or L5 Lagrangian points of stability. They are important solar-system fossils because they carry information on early Solar System formation, when collisions between bodies were more frequent. Discovery and study of terrestrial planet Trojans will help constrain models for the distribution of bodies and interactions in the inner Solar System. Since the discovery of the first outer planet Trojan in 1906, several thousand Jupiter Trojans have been found. Of the terrestrial planets there are four known Mars Trojans, and one Earth Trojan has been recently discovered. We present a new model that constrains optimal search areas, and imaging cadences for narrow and wide field survey telescopes including the \textit{Gaia} satellite for the most efficient use of telescope time to maximize the probability of detecting additional Earth Trojans.
\end{abstract}

\begin{keywords}
methods: numerical -- methods: observational --
minor planets, asteroids: general -- planets and satellites: general
-- celestial mechanics
\end{keywords}

\section{Introduction}

Trojan asteroids are minor planets that share the orbit of a planet about the Sun and librate around the L4 and L5 Lagrangian points of stability. The L4 and L5 points are $60\degr$ ahead and behind, respectively, the planet in its orbit. Trojans represent the solution to Lagrange's famous triangular problem and appear to be stable on long time-scales (100 Myr to 4.5 Gyr) \citep{pil99,sch05} in the N-body case of the Solar System. This raises the question whether the Trojans formed with the planets from the Solar nebula or were captured in the Lagrangian regions by gravitational effects. Studying the Trojans provides insight into the early evolution of the Solar
System.

Since the discovery of the first Trojan in 1906 \citep{nic61} several thousand more have been found in the orbit
of Jupiter. Of the terrestrial planets, four Trojans have been discovered in the orbit of Mars. There have been some attempts at searching for Earth Trojans (ET) \citep{dun83,whi98,con00}. Examination of a sky area of approximately 0.35 deg$^2$ by \citet{whi98} resulted in a crude upper limit on population estimates of $\sim3$ objects per square degree, down to $R=22.8$. The subsequent search by \citet{con00} covered $\sim9$ deg$^2$, down to $R=22$, with no ETs detected. Recent examination of data from the WISE satellite has resulted in the discovery of the first known ET \citep{con11}.

The earlier non-discovery may be attributed to a lack of observations targeting regions in which ETs could be found. For example, the programmes to search for Near-Earth Asteroids (NEA) generally employ observation strategies which survey regions having a higher
probability of containing potential Earth impactors. Typically these regions are near the plane of the ecliptic and do not specifically
target the entire region of stability for ETs. The Earth-Sun-ET geometry requires observations at low altitudes with limited observing time, restricting the ability to make multiple or follow-up observations. This geometry also results in low apparent motion relative to field stars as the orbit arc is at an oblique angle to an observer on Earth. Other reasons for non-detection include small population and orbital inclinations outside the plane of the ecliptic. Asteroids which could be Trojan candidates would not be flagged for further study by routine surveys because their apparent motions would not match the parameters of the survey for follow-up. 

Models \citep{mik90,tab00a} indicate ET stable orbits have some inclination to the plane of the ecliptic, in the range $10\degr$ to $45\degr$. Subsequent simulations \citep{mor02} predict $0.65\pm.012$ ETs with diameter $>$~1~km and $16.3\pm3.0$ asteroids with diameter $>$~100~m. The existence of the co-orbital asteroid 3753 Cruithne \citep{wie97}, and the recent discoveries of the horseshoe orbiter 2010~SO$_{16}$ \citep{chr11} and L4 Trojan 2010~TK$_7$ \citep{con11}, confirm the possibility of objects orbiting the Sun in 1:1 mean motion resonance with Earth. The question is how many, and what size are the bodies that share Earth's orbit as Trojan asteroids.

Assuming Trojans are approximately evenly distributed between the L4 and L5 regions then just one of these fields could be examined with the expectation that the other region would yield a similar result. However it should be considered that, given the short window of opportunity for making ground-based optical observations near morning (L4 region) and evening twilight (L5 region), both regions should be surveyed. With this in mind, it is important to find the optimal strategy to maximize sky coverage and probability of detection given those time constraints.

This paper employs a model probability distribution which we use to constrain optimal search areas and imaging cadences for efficient
use of telescope time while maximizing the probability of detecting ETs. 

\section{Model}

\begin{figure}
\includegraphics{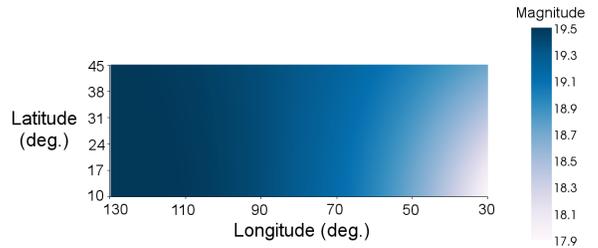}
\caption{\label{fig:figure1}Apparent magnitude of a 1~km Earth Trojan
by Heliocentric Longitude and Latitude. The figure shows variation
from $V=17.9$ at the nearest part of the field (to Earth) to $V=19.5$
at the farthest part.}
\end{figure}

\begin{figure}
\includegraphics{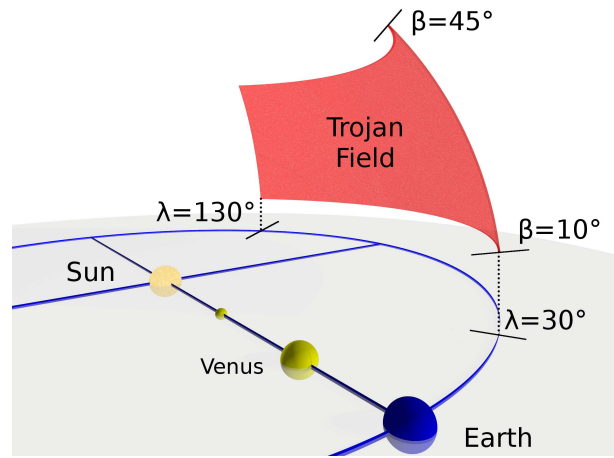}
\caption{\label{fig:figure2}Perspective illustration of Earth
Trojan (L4) target field. The field ranges from Heliocentric longitude
$(\lambda)$ $30\degr$ to $130\degr$ and latitude $(\beta)$ $10\degr$ to $45\degr$. A complementary
field exists in the trailing Lagrangian L5 region. This illustration
represents the field through which a body will pass during
its orbit.}
\end{figure}

Existing models for ETs provide estimates on populations but not composition. This is an important factor considering that albedo depends on composition. In the most likely case, ETs may be of S-type (silicaceous) similar to NEAs and inner Main Belt asteroids,
or they may be of C-type (carbonaceous). This influences the detection limit as S-types have an albedo $p_{v}=0.20$ and C-types have an albedo $p_{v}=0.057$ \citep{war09}. Calculations of absolute and apparent magnitudes for the classical L4 and L5 Lagrangian points, from \citet{ted05} and \citet{mor02}, are shown in Table \ref{tab:table1}, neglecting atmospheric extinction.  Assuming S-types as the dominant class in the inner Solar System the apparent magnitude for an ET of 1~km diameter varies between $V=17.9$ to $V=19.5$ (Fig. \ref{fig:figure1}), depending on the geometry across the field (Fig. \ref{fig:figure2}). An ET of 100~m diameter varies in magnitude between $V=22.9$ to $V=24.5$ across the field in the same fashion. 

\begin{table}
 \centering

\caption{Absolute and apparent magnitudes at the classical L4 and L5 Lagrangian points.\label{tab:table1}}

\begin{tabular}{lclcc}
\hline 
Type  & Albedo  & Diameter  & Abs. mag. $(H)$ & App. mag. $(V)$  \tabularnewline
\hline 
S-type & $0.20$ & 1.0~km & 17.37 & 18.9 \tabularnewline
  &   & 100~m & 22.36 & 23.9 \tabularnewline
\hline
C-type & $0.057$ & 1.0~km & 18.73 & 20.3 \tabularnewline
  &  & 100~m & 23.73 & 25.3 \tabularnewline
\hline 
\end{tabular}
\end{table}

\begin{figure}
\includegraphics{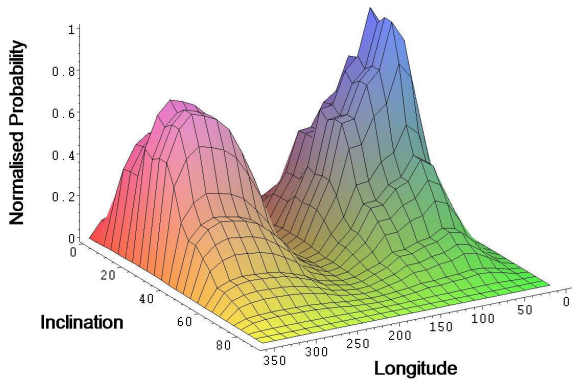}
\caption{\label{fig:figure3}Normalised probability
contour for Earth Trojan bodies by Inclination and Heliocentric Longitude
(degrees). The figure shows peak detection probabilities for longitudes
consistent with the classical Lagrangian points but that bodies, while
co-orbital with Earth, are unlikely to be co-planar.}
\end{figure}

A synthesis of the stable orbit inclination model \citep{mor02} and heliocentric longitude model \citep{tab00b} was used to identify probability regions where bodies are most likely to be (Fig. \ref{fig:figure3}). The ET fields (Fig. \ref{fig:figure2}) bounded by upper and lower inclination limits (FWHM) of $10\degr \la \beta \la 45\degr$
include $\sim74$ per cent of bodies \citep{mor02}. The heliocentric longitude limits (FWHM) of $30\degr \la \lambda \la 130\degr$
(L4 region) and $240\degr \la \lambda \la 340\degr$ (L5 region) include $\sim45$ per cent and $\sim40$ per cent of bodies respectively \citep{tab00b}. The regions bounded by these limits enclose $\sim63$ per cent of projected bodies. The recently-discovered ET, 2010~TK$_7$ \citep{con11}, has inclination $i=20.88\degr$ and mean longitude within the limits for the L4 region.

The heliocentric solid angle of each region is 0.931~sr (3056 deg$^2$). Calculation of the geocentric solid angle, necessary for Earth-based observations, requires a transformation from the heliocentric reference. A numerical integration is performed using the standard solid angle integral $\int\!\!\!\int_{S} \left(\mathbf{r} \cdot \mathbf{n}\right)/r^{3} dS$ (where $\mathbf{r}$ is the distance vector from the geocentre to the surface $dS$, $\mathbf{n}$ is the unit vector normal to that surface, and $r = \left| \mathbf{r} \right|$) to determine the solid angle subtended at points other than the centre. This integration enables calculation of sky area%
\footnote{The Python code used to calculate these solid angles is available from the author on request.%
}
for the heliocentric surface, from the geocentre for an Earth-based observer or for a space-based instrument such as the \textit{Gaia} satellite which will be located near the Earth's L2 Lagrangian point \citep{mig07}. The calculated geocentric solid angles are 0.3948~sr (1296 deg$^2$) and 0.4451~sr (1461 deg$^2$) for the L4 and L5 regions respectively. From the L2 Lagrangian point, for the \textit{Gaia} satellite, the solid angles are 0.3855~sr (1265 deg$^2$) and 0.4310~sr (1415 deg$^2$). The variation between regions is due to the slight differences in positions relative to Earth.

\subsection{Telescope surveys}

Having identified the ET fields we consider the ability of current and proposed wide-field telescopes to survey these regions. It is clear from the size of the regions ($>$1000 deg$^{2}$) that a widefield survey telescope would be needed for any attempt at surveying the entire region. However even existing telescopes with the widest fields of view would barely be able to survey the entire region once in a single day due to observing constraints imposed by the Earth-Sun-ET geometry.

We also consider the implications for a space-based telescope with the impending launch of the \textit{Gaia} satellite. \textit{Gaia} will not be subject
to the same constraints as ground-based telescopes. Although \textit{Gaia} has a relatively small FOV it will operate in a continuous scanning mode. It will image all of the sky down to a Solar elongation of $45\degr$ \citep{mig07}, but without the observational limitation of local horizon and airmass. However, this advantage is mitigated by its limiting magnitude of $V=20$.

\begin{figure}
\includegraphics{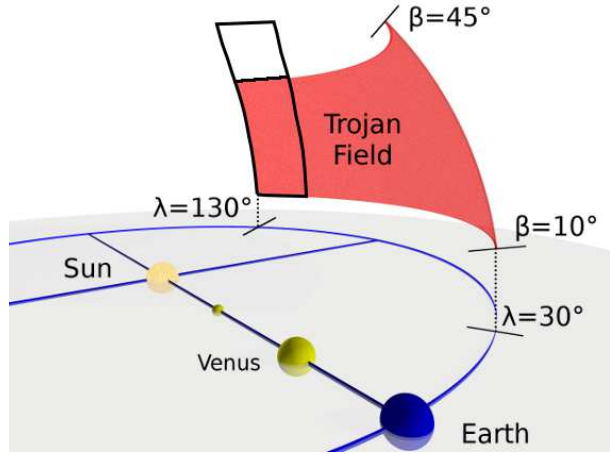}

\caption{\label{fig:figure4}Observing a defined range in Right
Ascension and Declination, with Earth's revolution about the
Sun, results in the entire field being imaged as it passes across
the observed region.}

\end{figure}

The continuous-scanning operation mode of \textit{Gaia} leads to the concept of a strategy of observing a swath of the region, a range in Right Ascension and Declination defining a particular sub-region. Observations can be made of that defined region, repeated twice per week, and making use of Earth's revolution about the Sun to progressively survey the target field. A $10\degr$-wide swath (Figure \ref{fig:figure4}) would be imaged in minutes by a survey telescope, as shown in Table \ref{tab:table2}. The region is redefined at monthly intervals to recommence the survey. Thus each month the target field passes through the observed region. In this fashion the entire field is resampled over a period of one year. This can be done at the beginning and end of the night, requiring minimal telescope time, before and after the primary science missions.

\begin{table*}
 \centering
 \begin{minipage}{180mm}

\caption{Comparison of different survey telescopes showing the required time
to survey the entire field compared to an optimal
strategy of observing a $10\degr$ swath and using Earth's revolution to
sweep across the field.\label{tab:table2}}

\begin{tabular}{llccccccl}
\hline 
Telescope  & Limiting mag.  & Exposure  & FOV  & \multicolumn{2}{c}{Entire field } & \multicolumn{2}{c}{$10\degr$ swath} & Instrument capabilities \tabularnewline
 &  &  &  & (FOVs) & (Time) & (FOVs) & (Time) &  \tabularnewline
\hline 
Catalina  & $V\sim20$ & 30~s & 8.0~deg$^2$ & 162/183 & 1.35h/1.5h & 12-18 & 6-9 min & \citep{dra09} \tabularnewline
PTF 1.2~m  & $R\sim20.6$  & 60~s  & 8.1~deg$^2$ & 160/180 & 2.7h/3h & 12-18 & 12-18 min & \citep{law09} \tabularnewline
Pan-STARRS & $R\sim24$ & 30~s & 7.0~deg$^2$ & 185/209 & 1.5h/1.7h & 13-20 & 7-10 min & \citep{jed07} \tabularnewline
LSST  & $r\sim24.7$ & 30~s & 9.6~deg$^2$ & 135/152 & 1.1h/1.3h & 10-15 & 5-8 min & \citep{jon09} \tabularnewline
\textit{Gaia} & $V\sim20$ & 39.6~s$^{\dagger}$ & 0.69\degr & 63/74 & 378h/444h$^{\ddagger}$ &  &  & \citep{mig07} \tabularnewline
\hline 
\end{tabular}

$\dagger$\textit{Gaia} will operate in a continuous scanning mode where the CCD array will be read out at a rate corresponding to the angular rotation rate of the satellite (6h period). The FOV value represents the number of rotations by \textit{Gaia}. \textit{Gaia's} specific precession parameters are not considered so values should be considered as representative.

$\ddagger$Approximate time to complete sufficient rotations to scan across the entire field. The actual time spent scanning within the field will be a fraction of this value.

\end{minipage}
\end{table*}

Observations of ETs are time-limited by the geometry of the Earth-Sun-ET positions. There exist specific constraints particular to the geographic location of a telescope. For example Northern Hemisphere telescopes will have access to the entire Northern part of the ET region described in Figure \ref{fig:figure2} while Southern Hemisphere telescopes will have access to only that portion visible above the local horizon. Some seasonal variation will occur with the change in relative orientations throughout the year, as noted in \citet{whi98}. The amount of variation will depend on geographic location.

By the end of evening twilight when it becomes possible to survey the L5 region, assuming (for convenience) the Sun's altitude is $-20\degr$, the classical Lagrangian point has an altitude of about $40\degr$. Hence atmospheric extinction on detection limits must be considered. Because there are many contributing factors to atmospheric extinction, reference to the ``average'' approximation given by \citet{gre92} suggests the extinction at $40\degr$ altitude is 0.44 mag and at $25\degr$ altitude is 0.66 mag for a site at sea level. 

The stable orbits of the ETs have been modelled to have an inclination range of $10\degr$ -- $45\degr$ \citep{mor02}. The heliocentric longitude ranges (FWHM) are $30\degr$ -- $130\degr$ and $240\degr$ -- $340\degr$ for L4 and L5 respectively \citep{tab00b}. Since the modelled orbits are inclined at some angle to the ecliptic the survey could be carried out in the upper or lower ecliptic latitudes $\left(\beta\right)$ at $10\degr\leq\beta\leq45\degr$ and $-10\degr\leq\beta\leq-45\degr$. By surveying fields in those latitudes the transverse motion per unit time is reduced. To an Earth-based observer an ET will appear to oscillate in an up-and-down motion about the plane of the ecliptic during its orbit about the Sun. The exhibited transverse motion will appear sinusoidal, with the apparent transverse motion reaching minimum at the extremes of its heliocentric latitude and reaching maximum when crossing the ecliptic plane.

It is impossible, a priori, to predict the direction of motion for the ET. Consequently it is not possible to apply a tracking offset for the apparent motion to increase the detection limit. A suitable alternative is to observe with the intention of imaging the ET at maximum heliocentric latitude when the apparent motion is at minimum. As it will then appear to be a slow-moving object, follow-up becomes difficult in the same night. However, repeat imaging can be performed on subsequent days rather than trying to acquire follow-up images during the same session. This approach also has the advantage of increasing the available observing time for a target field, if necessary.

While this delay between follow-up images introduces other variations from such things as changes in atmospheric conditions and seeing, this could be compensated for by image convolution. Some telescopes are implementing image processing systems designed specifically for asteroid detection (e.g. Pan-STARRS+MOPS -- Moving Object Processing System) where the asteroid is an apparently stationary transient because it is a very distant slow moving object \citep{jed07}. This method can be applied to nearer objects which are apparently slow moving as a result of the particular Earth-Sun-MP positions at the time of observation.

\section{Results}

Surveys of the entire field within the chosen limits are impractical on telescopes with small FOV but are reasonable on survey telescopes with sufficient FOV to accomplish the task in a single night, such as Catalina or the Large Synoptic Survey Telescope (LSST)%
\footnote{The LSST is still in the development phase (www.lsst.org).%
}. 
In these cases it may be possible to survey the regions defined by the inclination limits at intervals of approximately two months. This requires the fields to be surveyed twice within a few days. These surveys can be conducted at the beginning (L5 region) and end (L4 region) of the night. Although this is possible, it may be impractical as this occupies a significant amount of telescope time on those nights.

As the field spans $90\degr$ in longitude the intervals could nominally be three months. This increases the risk of missing low-inclination ETs completely through not having yet entered the field or of having just exited the field. It also increases the risk of detecting ETs shortly before exiting the field towards the ecliptic, possibly requiring follow-up observations in a less favourable orientation to the local horizon. By observing at intervals of two months a small amount of oversampling is introduced and the risk of missing an ET is reduced. On the assumption that only the Northern or Southern field could be surveyed depending on the location of the telescope, the program length would be 1 year, the length of the orbital period of a Trojan as it would pass through this field once per revolution. 

It is possible to conduct a survey in the ecliptic plane between latitude 0 and the lower limit of the field to search for Trojans as they cross the ecliptic. This reduces the necessary time per-session to survey the region as the ecliptic region is smaller than the ET field. The length of this program is reduced, requiring six months. Any Trojans detected in that six month period would be crossing either to the North or to the South. However the time saving per session is only about 30 per cent and requires more frequent sampling since, in this region, the ET will have a higher apparent motion. This could outweigh any benefit in the reduced per-session length as the total time requirement is not significantly different to that required to survey the entire field over the period of one year. 

Attempting to survey the entire field in a single session is challenging. The desirability of minimising the time per session is our key objective. Observing a swath of sky in a particular range of Right Ascension and Declination, and using Earth's revolution about the Sun to survey the field as it crosses the observed region, achieves this aim. This approach requires minimal time each session, i.e. two sessions per week during the course of one year. The observed region is redefined at monthly intervals to recommence sampling of the field. Some oversampling will occur as the field crosses the observed region. This results in nights lost due to adverse weather conditions becoming less critical to the overall program.

\section{Summary and Future Work}

Despite the thousands of known Jupiter Trojans, of the terrestrial planets a mere handful of Trojans have been discovered. There are only four known Mars Trojans, and the first ET has only been very recently discovered \citep{con11}. Simulations \citep{mor02} have predicted the existence of a number of ETs. The prospect of detecting these is limited by the small amount of time available per day due to the Earth-Sun-ET geometry. This implies that the conventional approach to detecting asteroids by repeated observations of a field must have cadence times in days rather than hours.

Surveys of the entire ET field are impractical due to the observational limits imposed by the geometry, and the many hours it would take to conduct such a survey. This paper has identified the region of highest probability for detection, in the inclination range of $10\degr$ -- $45\degr$ and heliocentric longitude range of $30\degr$ -- $130\degr$ (L4) and $240\degr$ -- $340\degr$ (L5).  The sky area for Earth-based observers has been determined using numerical integrations. A strategy has been proposed for observing a sub-region of the ET field and using Earth's revolution about the Sun to progressively survey the field as it crosses this sub-region. This approach takes only a few minutes per session, two days per week, and is readily achievable by a survey telescope before or after the primary science mission. While this method requires a program of continued observations for one year, the total time commitment for the program is a few tens of hours spread throughout that year. We note that the only known ET, 2010~TK$_7$, would pass through this region during its orbit.

The specific observing geometry of the \textit{Gaia} satellite and its position at Earth's L2 Lagrangian point will be examined in more detail in future work. Initial simulations for the detection of Trojans by \textit{Gaia} show promise (Mignard, pers. comm.). Results of detailed simulations will be reported with particular regard to the detection limits and observational mode of operation of \textit{Gaia}.

The different, and changing, geometry of the Earth-Mars Trojan relationship makes a detailed study more involved than the relatively static geometric condition of the ETs. As our nearest planetary neighbour with known Trojans, and the prediction that many more should exist in Mars' orbit \citep{tab99}, a similar modelling exercise is in progress. The peculiarities of the changing geometry and the implications for a search for additional Mars Trojans will be explored in greater depth.

The search area defined for ETs may also result in the detection of Inner Earth Objects (IEO), asteroids with orbits interior to Earth's orbit about the Sun, as they pass across the field. To date there have been nine IEOs discovered however the predicted number is $36\pm26$ with diameter $>$~1~km and $530\pm240$ with diameter $>$~250~m \citep{zav08}. Further work is required to examine the potential for discovery of these IEOs during an ET search.

\section*{Acknowledgments}
The authors would like to thank the anonymous referee whose comments and suggestions significantly improved the final version of this manuscript. M. Todd thanks the organisers of the \textit{Gaia} workshop (Pisa 2011) for providing a fertile environment for discussing \textit{Gaia} science. D.M. Coward is supported by an Australian Research Council Future Fellowship.

\label{lastpage}

\end{document}